# Study of Yu-Shiba-Rusinov bound states by tuning the electron density at the Fermi energy

Sang Yong Song[1,†], Yoon Sung Park[2,†], Yongchan Jeong[1], Min-Seok Kim[1], Ki-Seok Kim[2,*], and Jungpil Seo[1,*]

[1]*Department of Emerging Materials Science, DGIST, 333 Techno Jungang-daero, Hyeonpung-eup, Dalseong-gun, Daegu 42988, Korea*

[2]*Department of Physics, POSTECH, 77 Cheongam-ro, Nam-gu, Pohang 37673, Korea*

[†]The authors contributed equally to this work.

[*]Correspondence should be tkfkd@postech.ac.kr or jseo@dgist.ac.kr

## Abstract

**Magnetic atoms can break the Cooper pairs of superconductors, leading to the formation of Yu–Shiba–Rusinov (YSR) bound states inside superconducting gaps. Theory predicts that the YSR bound states can be controlled by tuning the electron density at the Fermi energy, but it has not been studied deeply. In this work, we studied the nature of YSR bound states in response to the potential scattering *U* by tuning the electron density at the Fermi energy. By comparing two systems, Mn-phthalocyanine molecules on Pb(111) and Co atoms on PbSe/Pb(111), we demonstrate that the sign of *U* can be unambiguously determined by varying the electron density at the Fermi energy. We also show that *U* competes with the exchange interaction *JS* in the formation of YSR bound states. Our work provides insights into the interactions between magnetic atoms and superconductors at a fundamental level.**

The local response of magnetic atoms to their environmental electronic structure is at the heart of technological applications in condensed matter systems. When magnetic adatoms on the surfaces of superconductors break Cooper pairs by interacting with quasiparticles through the exchange coupling $J$, excitations called Yu-Shiba-Rusinov (YSR) bound states appear inside the superconducting gap [1-3]. A simple explanation assuming the classical spin $S$ of the magnetic impurity shows that the energy of the bound states depends on the interaction strength $\rho JS$, where $\rho$ is the density of states (DOS) at the Fermi energy ($E_F$) and effectively controls the electron density at $E_F$ [4-14].

The relationship between YSR bound states and the exchange interaction ($JS$) has been theoretically proposed and demonstrated experimentally [15-20]. However, thus far, the impact of the electron density at $E_F$ on the YSR bound states has only been studied in the perturbative limit [21]. In this letter, we present experimental results demonstrating how the YSR bound states respond to strong variations in the electron density at $E_F$. To accomplish this, we have utilized the quantum confinement effect of Pb films on Ar cavities embedded in Pb(111) substrate [22]. By comparing two systems, Mn-phthalocyanine (MnPc) molecules on Pb(111) and Co atoms on PbSe/Pb(111), using scanning tunneling microscopy and spectroscopy (STM/STS), we show that not only $JS$ but $U$ plays a particularly important role in forming the YSR bound states on superconductors.

The experiments were performed using a home-built low-temperature STM. The temperature for measurements was 4.3 K. To maximize the spectrum resolution of the YSR bound states, we used a superconducting tip (*i.e.*, PtIr tip coated with Pb). The differential conductance ($dI/dV$) spectra were acquired using a standard lock-in technique. The measurement parameters and the details of the sample preparation are provided in Supplemental Material.

The Pb(111) single crystal was sputtered with 2 kV $Ar^+$ ions and subsequently annealed at 500 K, which introduced Ar cavities into the Pb(111) substrate (Fig. 1a). Figure 1b depicts the experimental concept where an Ar cavity is formed under the Pb surface. The Pb electrons are vertically confined between the Pb surface and the Ar cavity, forming quantum well states (QWS). The QWS depend on the thickness of Pb film above the cavity and generally vary from cavity to cavity. When one of the QWS is aligned at $E_F$, the electron density increases. Conversely, when $E_F$ is located between the QWS, the electrons are depleted. By depositing magnetic impurities on the Ar cavities, we can probe the superconducting response to the magnetic impurities depending on the electron density at $E_F$. In applying our idea to real systems, we first investigated MnPc molecules grown on Pb(111). The YSR bound states induced by MnPc molecules on Pb(111) have been well-studied [16,17,23], and thus, investigation of this system provides general comprehension regarding the response of YSR bound states to the electron density variations.

Figure 2a shows the topography of self-assembled MnPc molecules on Pb(111). Figure 2b displays representative QWS induced by Ar cavities. The QWS correspond to van Hove singularity (vHs) peaks resulting from the vertical confinement of the 3-D electronic states of Pb. For most Ar cavities, $E_F$ is in between the QWS, leading to suppressed electron density. This is to minimize the total energy of the system, and has been reported in Pb films on Si(111) and Cu(111) [24,25]. Due to the suppressed electron density at $E_F$, Pb films on Ar cavities are imaged with dark contrast when scanned with low bias voltage ($V_{bias}$). Therefore, the reduced contrast around molecules in Fig. 2a indicates the presence of an Ar cavity under the Pb surface (indicated by the hexagon).

Figure 2c shows the YSR bound states measured on the MnPc molecule marked by the blue arrow in Fig. 2a. Because we used a superconducting tip to probe the YSR bound states, the superconducting coherent peaks are positioned at the energy ($E = eV_{bias}$) of $E = -\Delta_S - \Delta_T$ and $\Delta_S + \Delta_T$, where $e$ is the electron charge, $\Delta_S$ and $\Delta_T$ are superconducting gaps of the Pb substrate and the tip, respectively. The shaded area between $-\Delta_T$ and $\Delta_T$ in Fig. 2c represents the tip gap. The blue and red vertical arrows in Fig. 2c correspond to the YSR excitations at negative and positive energies, respectively. They are symmetric in energy with respect to $E_F$, but the intensities are asymmetric due to the potential scattering $U$ of Mn [7]. As a convention, we define the energy ($\varepsilon$) of the YSR bound states using a peak with stronger intensity. To exclude the tip gap in determining $\varepsilon$, we measured $\varepsilon$ from $\Delta_T$ or $-\Delta_T$ depending on the sign of $\varepsilon$ (see the graphical definition of $\varepsilon$ in Fig. 2c).

Figure 2d exhibits the *dI/dV* map measured at $V_{bias}$ = -1.3 mV. The pattern shows the spatial variations in YSR bound states due to the *JS* modulation of MnPc molecules on Pb(111) [17]. To compare the YSR bound states of MnPc molecules on the Ar cavity and Pb bulk, two horizontal dashed lines are selected in Fig. 2d, where one crosses the cavity and the other does not. They are good candidates for comparison because the pattern of the *dI/dV* map is equivalent along these lines. Figures 2e and 2f show spatially resolved *dI/dV* spectra along the dashed lines. Compared with the MnPc molecules on the Pb bulk (Fig. 2f), the YSR bound states of the MnPc molecules on the Ar cavity are pushed toward the gap edge as indicated by the yellow arrow in Fig. 2e. Figure 2g displays the *dI/dV* spectra for the MnPc molecule marked with ‘3’ in Figs. 2e and 2f. Using MnPc molecules in equivalent moiré adsorption sites, we quantified the YSR energy $\varepsilon$ for 79 MnPc molecules on the Pb bulk and 14 MnPc molecules on 6 different Ar cavities (Fig. 2h and Supplemental Material). The YSR energies of the molecules on Pb bulk are scattered within the shaded area in Fig. 2h. In contrast, the YSR energies on the Ar cavities are shifted toward $+\Delta_S$ as a consequence of the electron depletion on the Ar cavities. This is consistent with the theoretical prediction, as the YSR energy is given by [3,7,26]

$$\varepsilon = \Delta_S \frac{1+(\rho U)^2-(\rho JS)^2}{\sqrt{[1+(\rho U)^2-(\rho JS)^2]^2+4(\rho JS)^2}}. \quad \dots (1)$$

The YSR energy moves to gap edge when $\rho$ is decreased. Similar results were also reported when *J* was reduced by pulling the magnetic molecules from Pb(111) using a tip [18,19].

We then further investigated YSR bound states induced by Co atoms on 1-monolayer (ML) PbSe grown on Pb(111). The superconductivity of PbSe is proximity-induced from Pb(111) [27]. Figure 3a shows a topographic image of Co atoms on PbSe/Pb(111). The Ar cavities are imaged with dark contrast, ensuring that the electron density is diminished at $E_F$. The inset shows the image of a Co atom placed on an Ar cavity. The QWS measured on this cavity are displayed in Fig. 3b. Figure 3c shows YSR bound states of a Co atom placed on the bare PbSe surface. The YSR energy is located at the negative bias according to our convention. The YSR bound states measured on the Co atom on the cavity are shown in Fig. 3d. In contrast to the MnPc molecules on Pb(111), the YSR energy is pushed toward $-\Delta_S$, which is indicated by the red arrow in Fig. 3d. Figure 3e displays the statistical distribution of YSR energies of 67 Co atoms on the bulk and 24 Co atoms on Ar cavities. By assuming that $\rho$ is inversely proportional to the QWS gap [25], the YSR energy shifts toward $-\Delta_S$ as $\rho$ decreases (Fig. 3f), the tendency of which is contrary to that of the MnPc molecules on Pb(111).

To understand the different trends regarding YSR energy shift, we formulated the YSR bound states by solving the Schrödinger equation (Supplemental Material). Figures 3g and 3h show the calculated YSR bound states using the parameter $U$ = 2.5 and $U$ = -2.5, respectively. For both calculations, $JS$ = 1.7, $\rho$ = 1 and $\Delta_S$ =1 are used. When $U > 0$, the YSR excitation at positive energy ($|\psi^+|^2$) is stronger than the YSR excitation at negative energy ($|\psi^-|^2$) in a weak coupling regime (*i.e.*, $\rho JS < w_c = \sqrt{1+(\rho U)^2}$) (Fig. 3g). When $U < 0$, the intensities are reversed, and $|\psi^-|^2$ is stronger than $|\psi^+|^2$ in the weak coupling regime (Fig. 3h). This is because the symmetry breaking of the particle-hole intensities depends on the sign of $U$, which is shown theoretically by Salkola *et al.* [26] However, its experimental proof is lacking because there is an ambiguity in determining whether a magnetic impurity on a system is weak or strong for $\rho JS$. For instance, in both cases of (a) $U > 0$ with $\rho JS < w_c$ and (b) $U < 0$ with $\rho JS > w_c$, the spectral weight of $|\psi^+|^2$ is larger than $|\psi^-|^2$, and they are indistinguishable [26].

In our experiments, we push the systems toward the weak coupling regime of $\rho JS$ by reducing the electron density at $E_F$, which removes the aforementioned ambiguity. Therefore, we determine that $U > 0$ for the MnPc molecules on Pb(111) because the YSR energy shifts toward $+\Delta_S$ as $\rho JS$ decreases. Similarly, we find that $U < 0$ for the Co atoms on PbSe because the YSR energy shifts toward $-\Delta_S$ as $\rho JS$ decreases. Since the YSR intensities at positive and negative energies are switched to each other by the sign of $U$, the YSR energy satisfies the condition $\varepsilon(-U) = -\varepsilon(U)$.

Now we discuss the sign of $J$ for the Co atoms on PbSe/Pb(111). When the measurement temperature was raised up to 7.7 K, Kondo resonance peaks appear for the Co atoms (Supplemental Material). This suggests that the exchange coupling between the magnetic moment and host electrons is antiferromagnetic ($J > 0$ for our Hamiltonian in Supplemental Material) [7,15,28,29]. Otherwise, the spin would have been effectively decoupled from the system as known in the numerical renormalization group calculations [30,31].

Although the electron density at $E_F$ is diminished for most Ar cavities, we occasionally find Ar cavities that show enhanced electron density at $E_F$. Such a cavity appears as a bright island in the topographic image (Fig. 4a). The *dI/dV* spectrum measured on the bright island clearly shows that one of the QWS is at $E_F$ (Fig. 4b). Figure 4c shows a zoomed-in image of the bright island, revealing that there is a Co atom on the island. Figure 4d shows the YSR bound states measured on the Co atom. Remarkably, the spatial oscillations of the YSR bound states are observed on this cavity (Figs. 4e and 4f) [14,32]. We attribute these apparent oscillations of the YSR bound states to the focusing effect of Fermi surface [33-36], which is caused by the vHs peaks. As the $k_z$ states are restricted through the geometric confinement, $k_x$ and $k_y$ states of similar magnitude are populated near the allowed $k_z$ states, which makes the lateral decay of the YSR bound states less predominant. The YSR wavefunctions in space are given by (Supplemental Material and ref. [3])

$$\psi_\uparrow(\mathrm{r}), \psi_\downarrow(\mathrm{r}) \propto \frac{sin(k_F r - \delta^\pm)}{r} exp(-r|\sin(\delta^+ - \delta^-)|/\xi), \ldots (2)$$

where $k_F$ is the Fermi wavevector and $\xi$ is the coherence length of the superconductor, $\delta^+$ and $\delta^-$ are the scattering phases for the spin-up part and spin-down part of the YSR bound states, respectively. In the weak coupling regime with $J > 0$, $|\psi_\uparrow(\mathrm{r})|^2$ and $|\psi_\downarrow(\mathrm{r})|^2$ are assigned to $|\psi^-(\mathrm{r})|^2$ and $|\psi^+(\mathrm{r})|^2$, respectively (Supplemental Material). The YSR energy shown in Eq. (1) can be rewritten as $\varepsilon = \mathrm{sign}(U)\Delta_\mathrm{S}\cos(\delta^+ - \delta^-)$ in terms of $\tan(\delta^+) = \rho(U + JS)$ and $\tan(\delta^-) = \rho(U - JS)$, considering the sign of $U$. Figures 4e and 4f show $|\psi^-(\mathrm{r})|^2$ and $|\psi^+(\mathrm{r})|^2$. Figure 4g shows the line profiles across the Co atom in Figs. 4e and 4f. We found that $k_F$ is approximately 2.9 $\mathrm{nm}^{-1}$ and the phase difference ($\delta^+$ - $\delta^-$) is approximately $0.3\pi$. This translates into YSR energy of $\varepsilon = -0.59\Delta_S$, which agrees with the measured value of $\varepsilon = -0.61\Delta_S$ in Fig. 4d.

Finally, we discuss the shift of YSR energy when the electron density is high at $E_F$. In Fig. 3f, the YSR energy shifts toward $-\Delta_S$ as the electron density decreases. Therefore, it is expected that the YSR energy will move toward $+\Delta_S$ when the electron density increases. However, we observed the YSR energy shifted again toward $-\Delta_S$ for some Co atoms on the Ar cavities with enhanced electron density. The example includes the Co atom represented in Fig. 4d. We provide a plausible scenario to explain this observation in terms of competition between $U$ and $JS$. When $|U| < |JS|$, it is found that $\delta^+ = \tan^{-1}(\rho(U + JS)) > 0$ and $\delta^- = \tan^{-1}(\rho(U - JS)) < 0$. Recall that $U < 0$ and $JS > 0$ for the Co atoms on PbSe/Pb(111). By increasing the electron density at $E_F$, $\delta^+$ and $\delta^-$ converge to $\pi/2$ and $-\pi/2$, respectively (the inset of Fig. 4h). The value of $\delta^+ - \delta^-$ ranges from 0 ($\rho \to 0$) to $\pi$ ($\rho \to \infty$), driving $\varepsilon$ to vary from $-\Delta_S$ to $\Delta_S$ (Fig. 4h). At $\delta^+ - \delta^- = \pi/2$, the quantum phase transition occurs and the ground state is switched from the spin-doublet state to the spin-singlet state [5,7,17,26]. In contrast, the signs of $\delta^+$ and $\delta^-$ are the same when $|U| > |JS|$ (*i.e.*, $\delta^+ = \tan^{-1}(\rho(U + JS)) < 0$ and $\delta^- = \tan^{-1}(\rho(U - JS)) < 0$). Both negative $\delta^+$ and $\delta^-$ then converge to $-\pi/2$ by increasing the electron density, enabling the value of $\delta^+ - \delta^-$ to approach 0 (the inset of Fig. 4i). Therefore, the YSR energy shifts toward $-\Delta_S$ although the electron density is increased at $E_F$ (Fig. 4i). This is consistent with the theoretical note that the spin-doublet state is thermodynamically stable when $|U| > |JS|$ [26]. We simulated the YSR oscillations using Eq. (2) and found that the results with $\delta^+ = -0.3 \times \pi/2$ and $\delta^- = -0.9 \times \pi/2$ agree well with the experimental data in Figs. 4e and 4f (Supplemental Material). The results are self-consistent with $U < 0$ and $J > 0$, and the signs of the phase shifts support that the system indeed falls into the criteria of $|U| > |JS|$.

In summary, we have studied the response of YSR bound states to electron density variations at $E_F$. By decreasing the electron density at $E_F$, we have determined $U > 0$ for the MnPc molecules on Pb(111) and $U < 0$ for the Co atoms on PbSe/Pb(111). By increasing the electron density at $E_F$, we reveal a competition between potential scattering $U$ and exchange interaction $JS$. We have shown that the ground state of the system can remain in a spin-doublet state when $U$ is stronger than $JS$. Our experiment explicitly shows that not only the exchange interaction $JS$ but the potential scattering $U$ is important in the formation of YSR bound states. This study deepens our understanding of how magnetic atoms interact with superconductors, and should be useful for designing future topological objects using magnetic atoms on superconductors.

## Acknowledgements

We are grateful to M.-S. Choi for the fruitful discussion. This work was supported by the National Research Foundation of Korea (NRF) grant funded by the Korea government (No. 2020R1A2C2102838 and No. 2011-0030046) and CoE program (19-CoE-NT-01) through DGIST R&D Program.

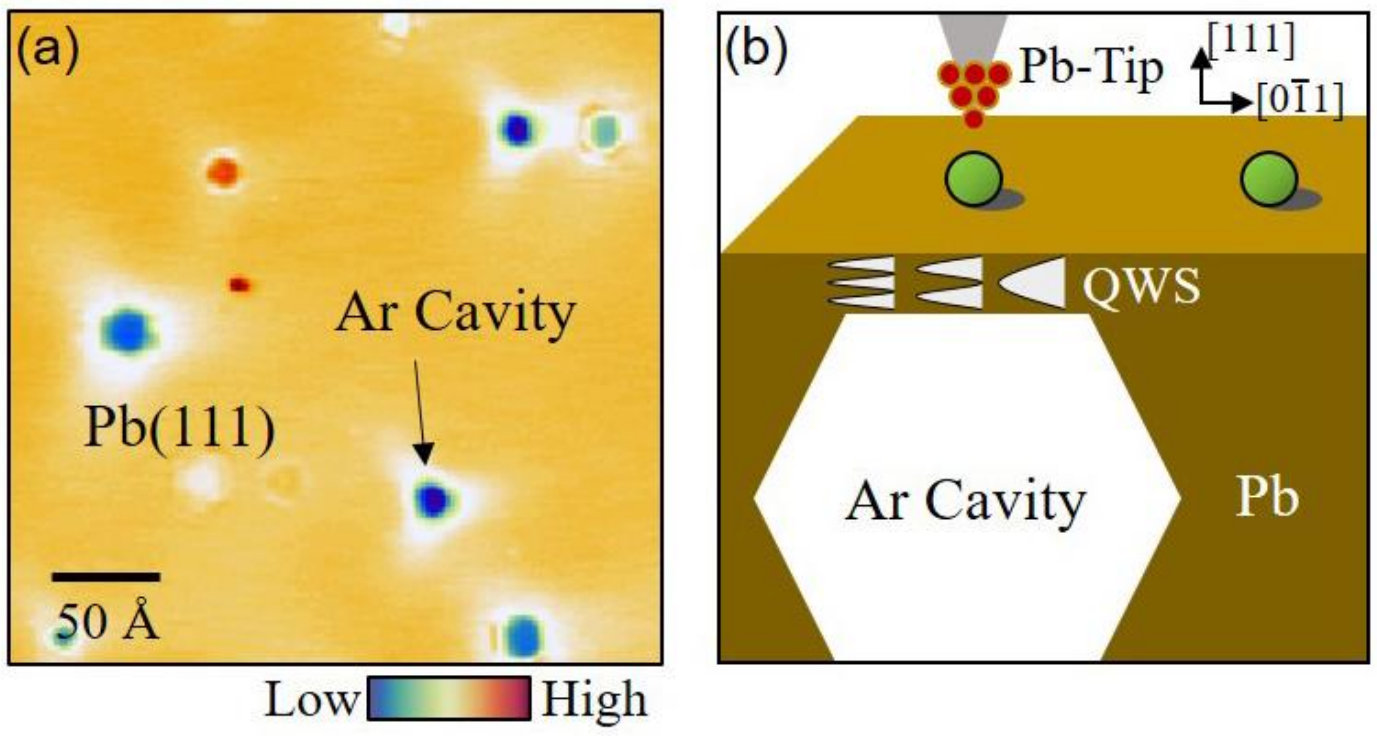

Figure 1. (a) Ar cavities are formed underneath the Pb surface. (b) The experimental concept. The QWS by the Ar cavity change the electron density at $E_F$. The green balls represent magnetic impurities on and off the Ar cavity.

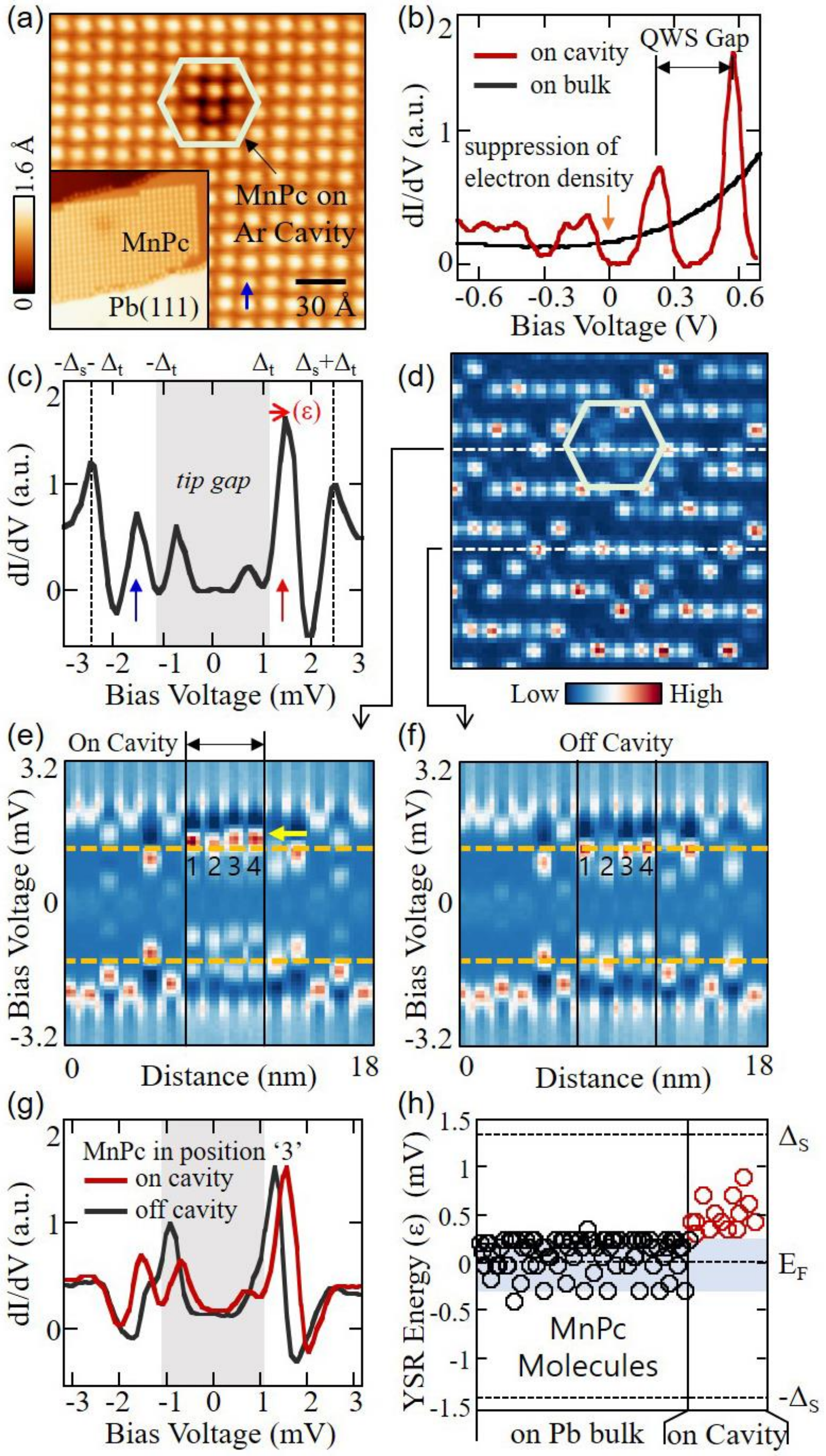

Figure 2. (a) The self-assembled MnPc molecules on Pb(111). (b) The QWS measured on an Ar cavity. (c) The *dI/dV* spectrum measured on the MnPc molecule marked by the blue arrow in (a). (d) The *dI/dV* map measured at $V_{bias}$ = -1.3 mV. (e,f) The *dI/dV* line spectra measured along the dashed lines in (d). The horizontal dashed lines are provided as a guide for energy comparison of the YSR bound states. (g) The *dI/dV* spectra on/off the cavity for the MnPc molecule in position '3' in (e) and (f). (h) The statistical distribution of YSR energy of MnPc molecules on/off the Ar cavities.

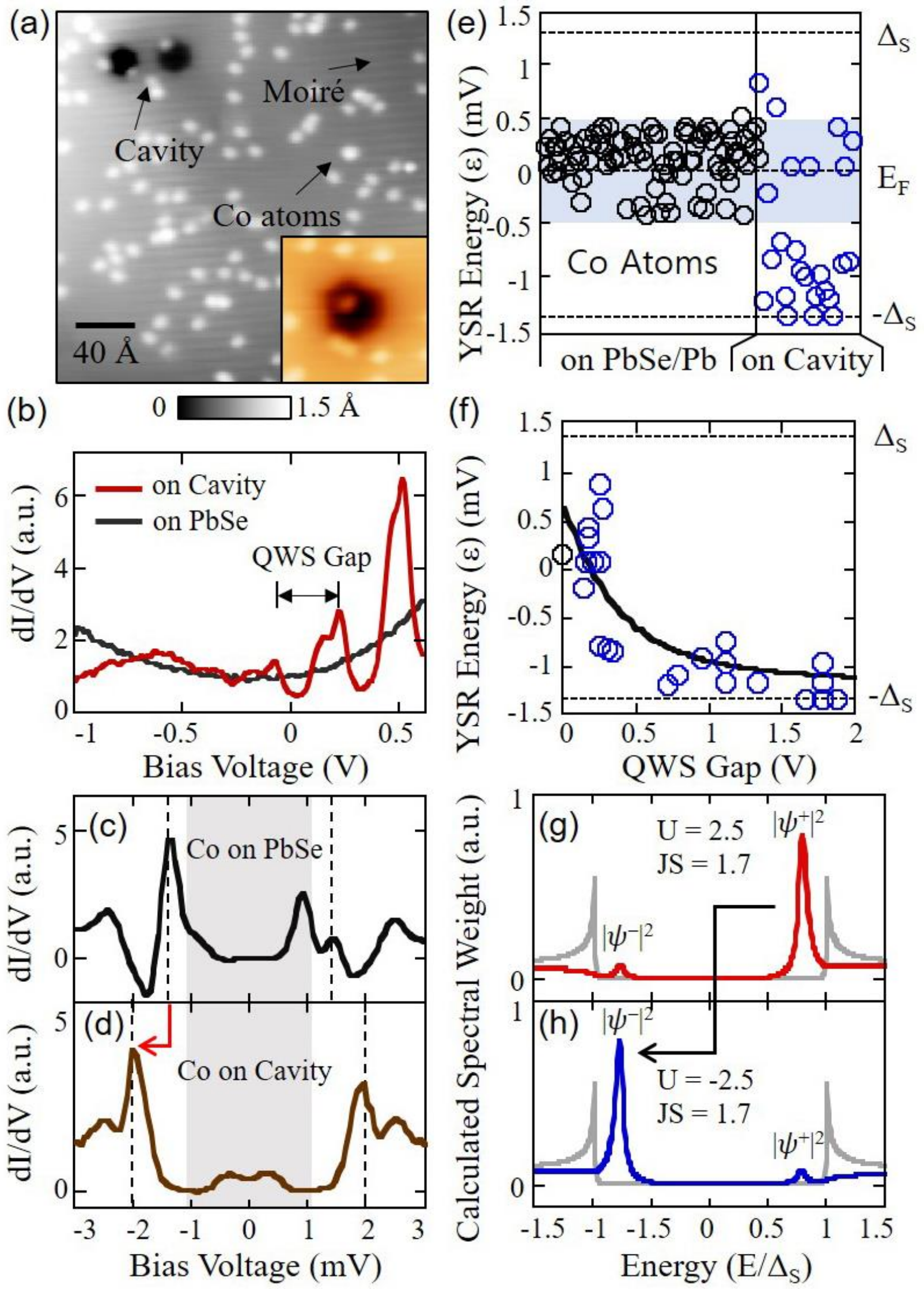

Figure 3. (a) The Co atoms on 1-ML PbSe grown on Pb(111). (b) The QWS measured on the cavity in the inset of (a). (c) The *dI/dV* spectrum measured on a Co atom on the PbSe. The vertical dashed lines indicate the positions of the YSR bound states. (d) The *dI/dV* spectrum measured on the Co atom on the Ar cavity in the inset of (a). (e) The statistical distribution of YSR energy of Co atoms on/off the Ar cavities. (f) Dependence of the YSR energy on the QWS gap. Solid line is an eye-guide plotted based on Eq. (1) in the text. (g) Simulation of the YSR bound states using $U$ = 2.5, $JS$ = 1.7, $\rho$ = 1 and $\Delta_S$ = 1. The background is calculated with $JS$ = 0. (h) When calculated using $U$ = -2.5, the YSR intensities are switched, as indicated by the black arrow.

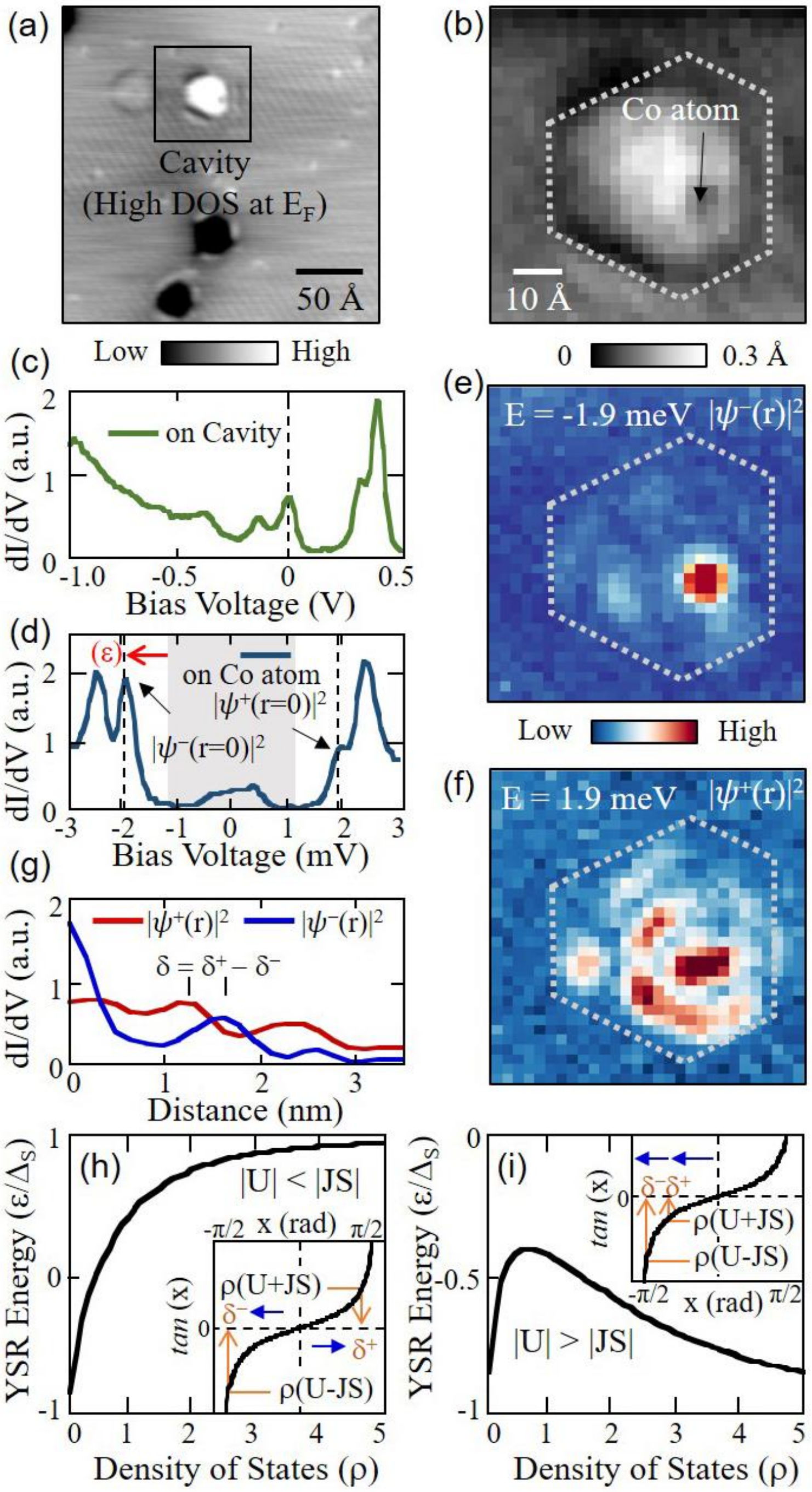

Figure 4. (a) The Ar cavity with enhanced electron density at $E_F$. (b) The *dI/dV* spectrum measured on the cavity. (c) Co atom located on the cavity. (d) The YSR bound states measured on the Co atom. The YSR energy ε is -0.8 mV. (e,f) The *dI/dV* maps for the YSR excitation at the negative and positive bias voltages. (g) The line profiles across the Co atom in (e) and (f). (h) (Calculation) The YSR energy depending on the electron density (ρ) at $E_F$ when $|U| < |JS|$. $U$ = -2.5, $JS$ = 3.2 and $\Delta_S$ = 1 were used in the calculation. The blue arrows indicate the direction of change for $\delta^+$ and $\delta^-$ when ρ is increasing. (g) (Calculation) The YSR energy when $|U| > |JS|$. $U$ = -3.5, $JS$ = 3.2 and $\Delta_S$ = 1 were used.

# Supplemental Material
# Study of Yu-Shiba-Rusinov bound states by tuning the electron density at the Fermi energy

**Sang Yong Song[1,†], Yun Sung Park[2,†], Yongchan Jeong[1], Min-Seok Kim[1], Ki-Seok Kim[2,*], Jungpil Seo[1,*]**
[1]Department of Emerging Materials Science, DGIST, 333 Techno Jungang-daero, Hyeonpung-eup, Dalseong-gun, Daegu 42988, Korea
[2]Department of Physics, POSTECH, 77 Cheongam-ro, Nam-gu, Pohang 37673, Korea
[†]The authors contributed equally to this work.
[*]Correspondence should be tkfkd@postech.ac.kr or jseo@dgist.ac.kr

# Contents

## 1 Preparation of Samples: MnPc Molecules on Pb(111) and Co Atoms on PbSe/Pb(111)

The Pb(111) single crystal was cleaned by the repeated sputtering (2 kV $Ar^{+}$) and annealing (500 K) cycles. This process introduced Ar cavities under the atomically flat Pb(111) surface. For the self-assembly of MnPc molecules on Pb(111), we heated the molecules up to 550 K using a Knudsen cell and evaporated onto Pb(111) for 8 minutes. The Pb(111) was kept at room temperature during the deposition. The post-annealing was followed at 330 K for 5 minutes. In growing 1-ML PbSe on Pb(111), we evaporated the selenium atoms onto Pb(111) for 3 minutes at the flux ratio of 1 Å/s. The substrate temperature was kept at 480 K during the deposition. The grown PbSe showed a rectangular crystal structure which is distinguished from the Pb crystal structure (Figure S1). The atomic spacing of the PbSe is about 6.0 Å, which corresponds to the lattice constant of PbSe(001). Moiré patterns were visible in the 45° direction from the PbSe lattice due to the lattice mismatch between PbSe(001) and Pb(111). The Co atoms were evaporated onto the PbSe/Pb(111) using a home-made e-beam evaporator.

A PtIr tip was sharpened by standard field emission processes with $V$ = 1.5 kV and $I$ = 10 $\mu$A. The tip was then coated with Pb by indentation into the Pb surface. The quality of the superconducting tip was checked by measuring $dI/dV$ spectra on the Pb surface.

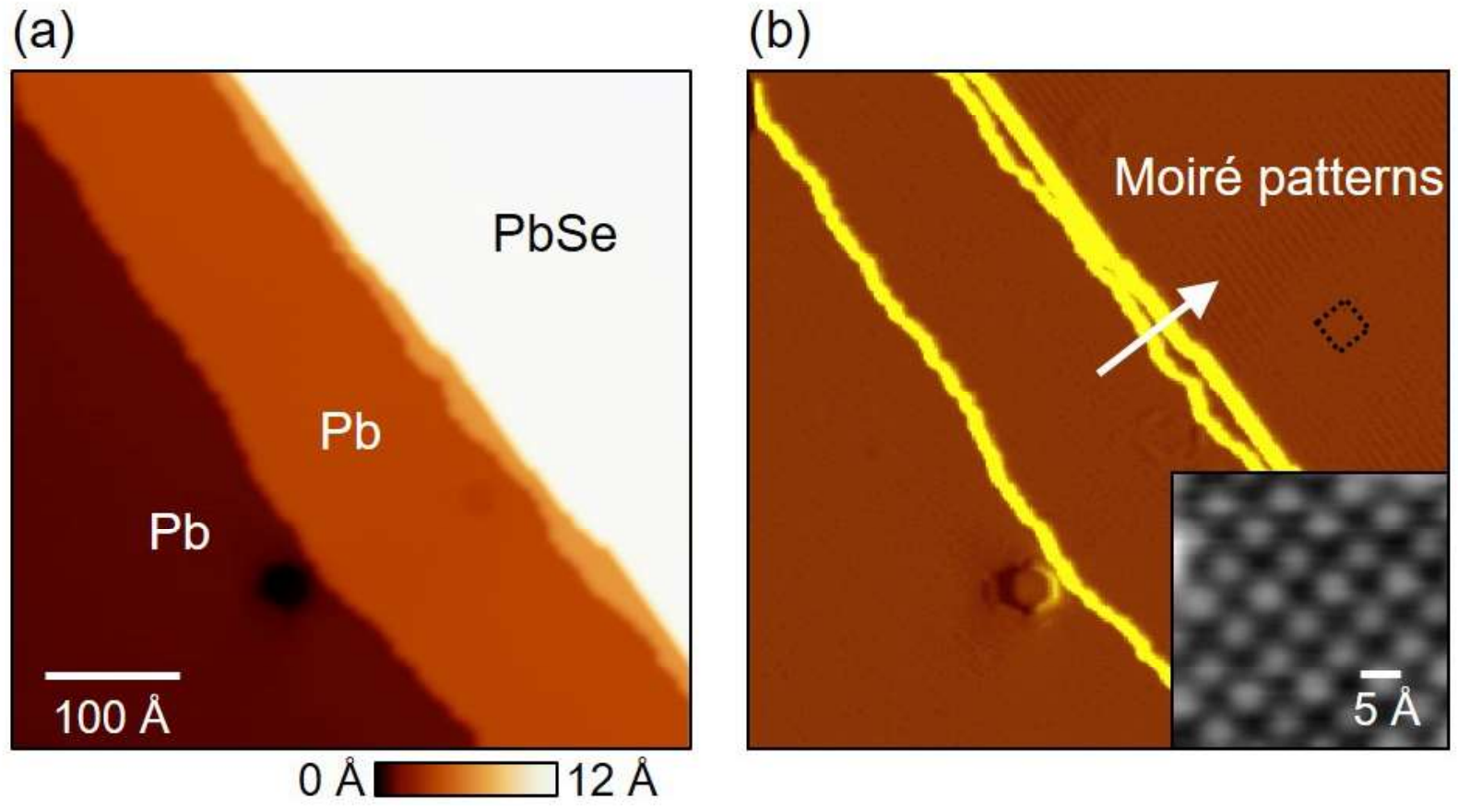

Figure S1. (a) The image shows 1-ML PbSe grown on Pb(111) substrate. (b) The derivative image of (a). When the area of the dashed box is enlarged, the atomic lattice of PbSe is revealed (the inset). The moiré patterns are observed on the PbSe due to the lattice mismatch between the PbSe and Pb substrate.

## 2 Characterization of YSR Energies of MnPc Molecules on and off the Ar Cavities

To compare between the YSR energies of the MnPc molecules on bulk and on cavities, we first defined the equivalent adsorption sites of the molecules on Pb(111). Figure S2 shows the $dI/dV$ map measured at $V_{bias}$ = -1.3 mV. The orange boxes in Fig. S2(b) denote such equivalent adsorption sites. The YSR bound states of the molecules in these orange boxes are deep inside the superconducting gap, therefore, we can easily distinguish the shift of the YSR energies responding to the cavities. By inspecting the spectra measured along each line of the $dI/dV$ map, we read out the YSR energies of the molecules (Fig. S3). When the orange box happened to coincide with the Ar cavities, the effect of the cavities was examined in altering the YSR energies (Figs. S4 and S5).

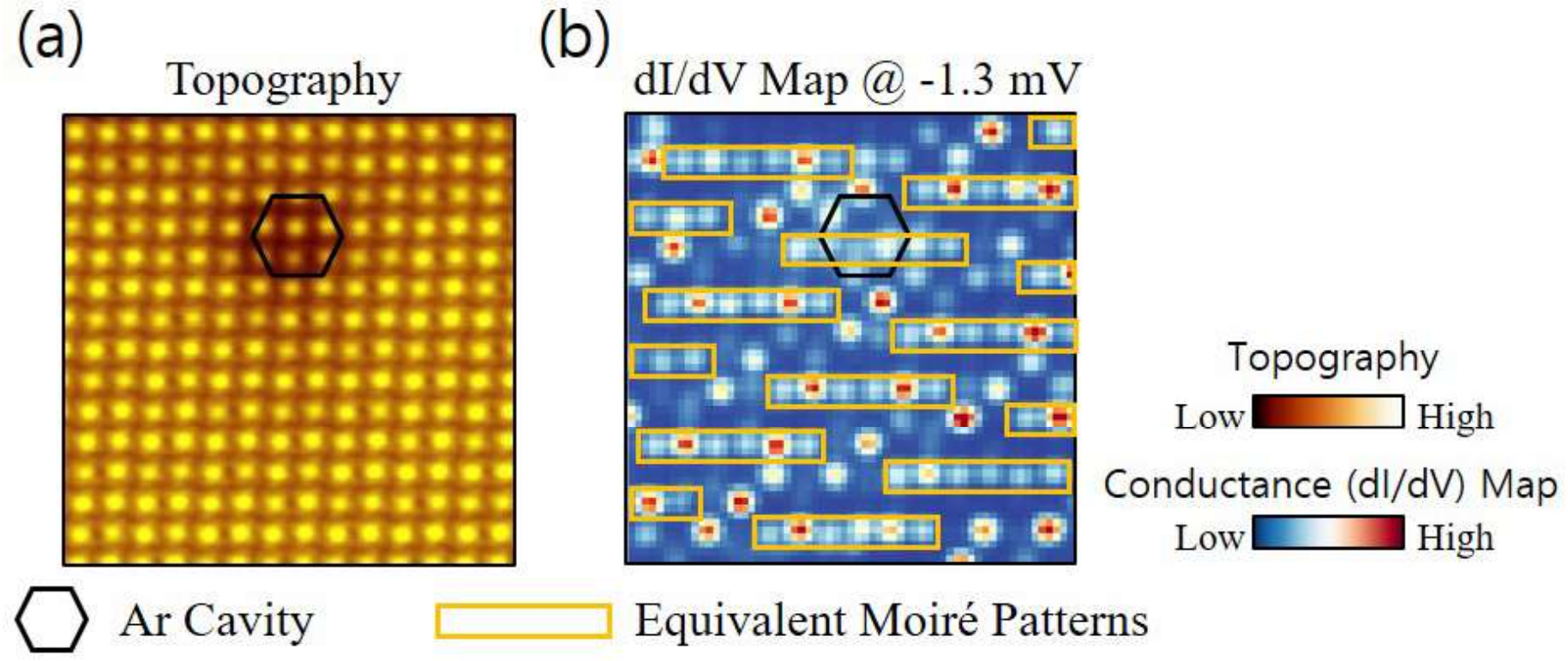

Figure S2. (a) Topography of MnPc molecules on Pb(111). $V_{bias}$ = -100 mV and $I$ = 50 pA. (b) The $dI/dV$ map at $V_{bias}$ = -1.3 mV. Equivalent moiré patterns are represented by the orange boxes.

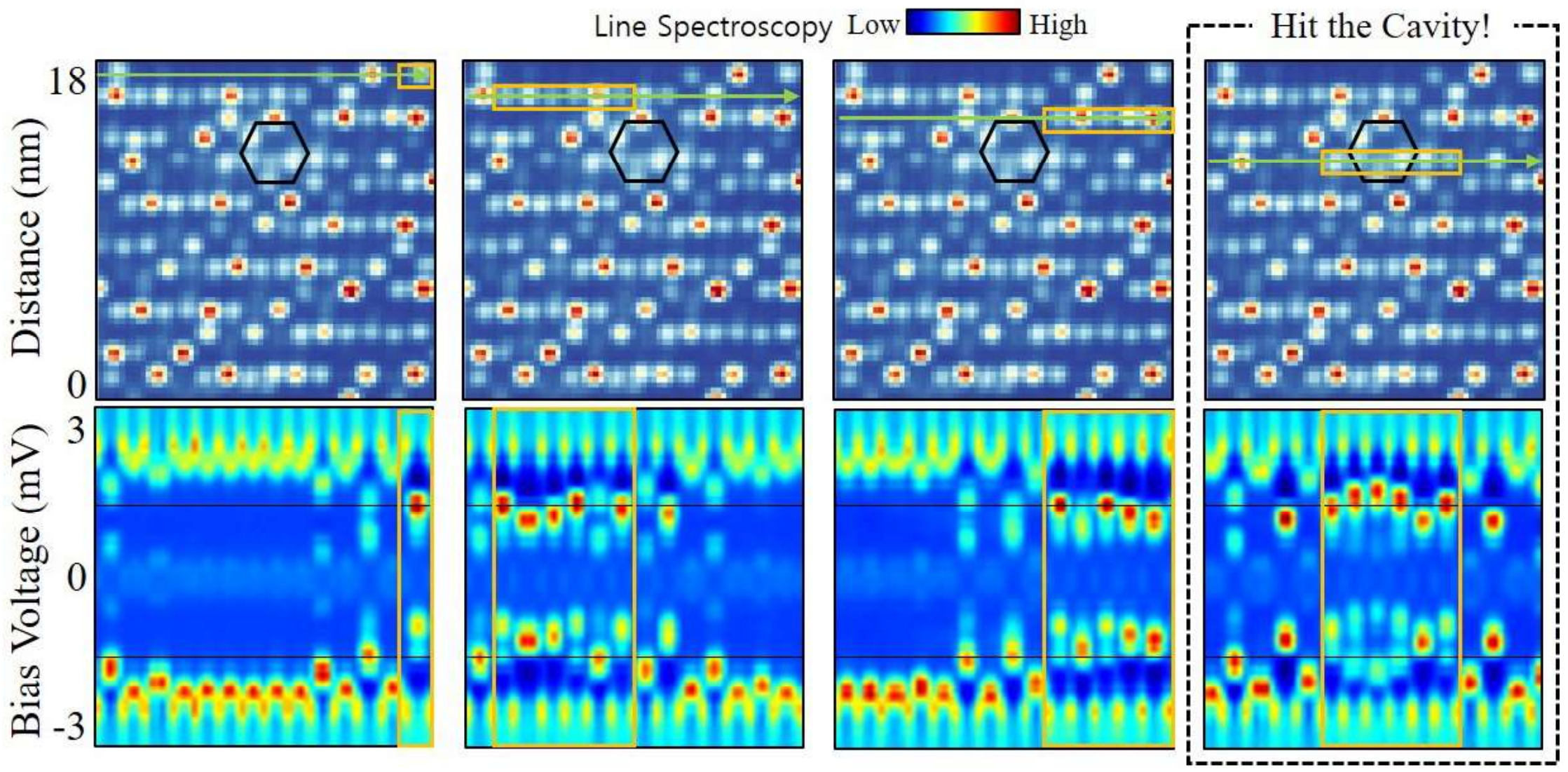

(continue on the next page)

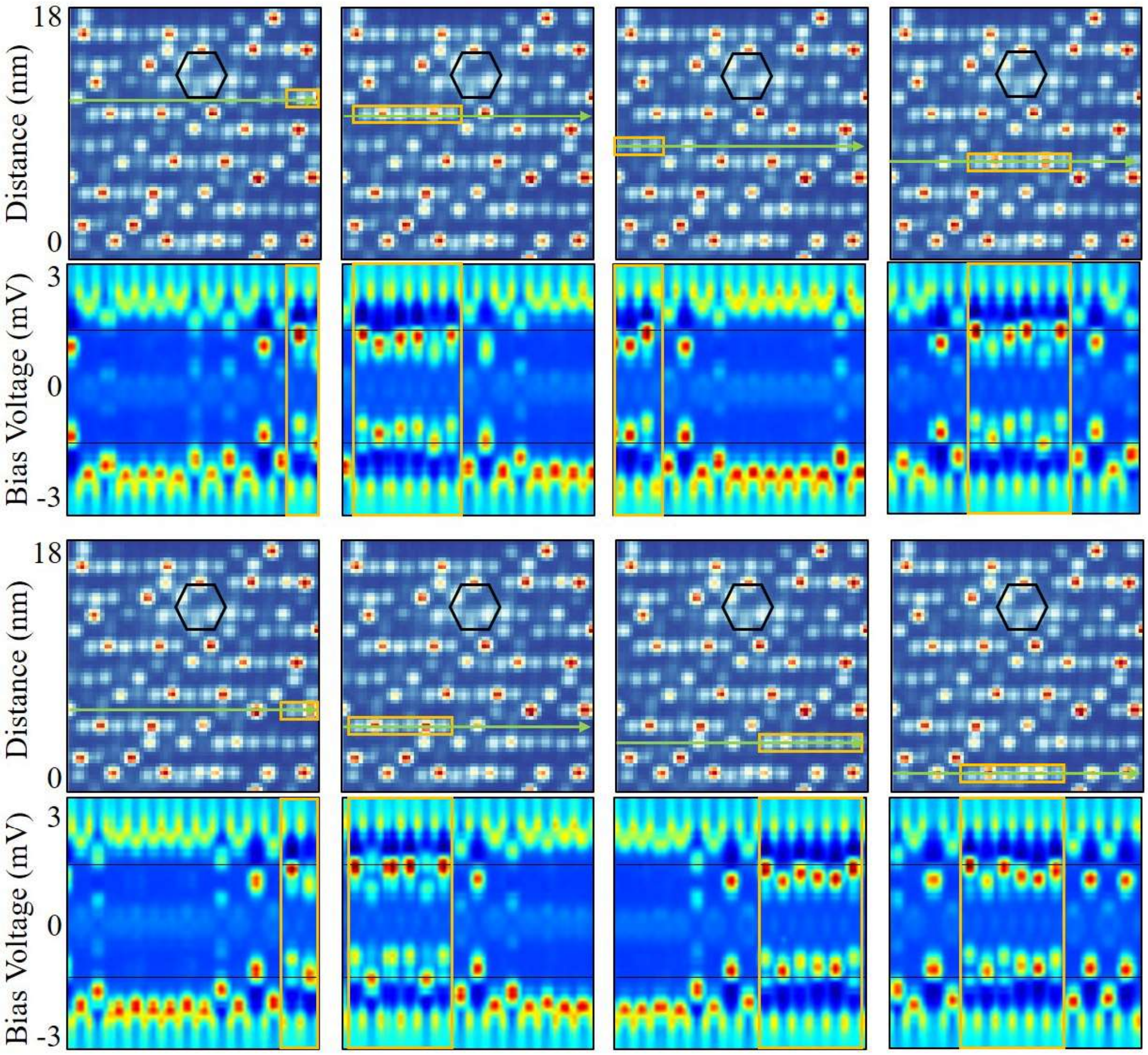

Figure S3. List of $dI/dV$ line spectra. Only the YSR bound states in the orange boxes are selected as data for analysis. When the orange box coincides with an Ar cavity (see the $dI/dV$ line spectra in the dashed box), the YSR energy shift can be examined in response to the DOS variations.

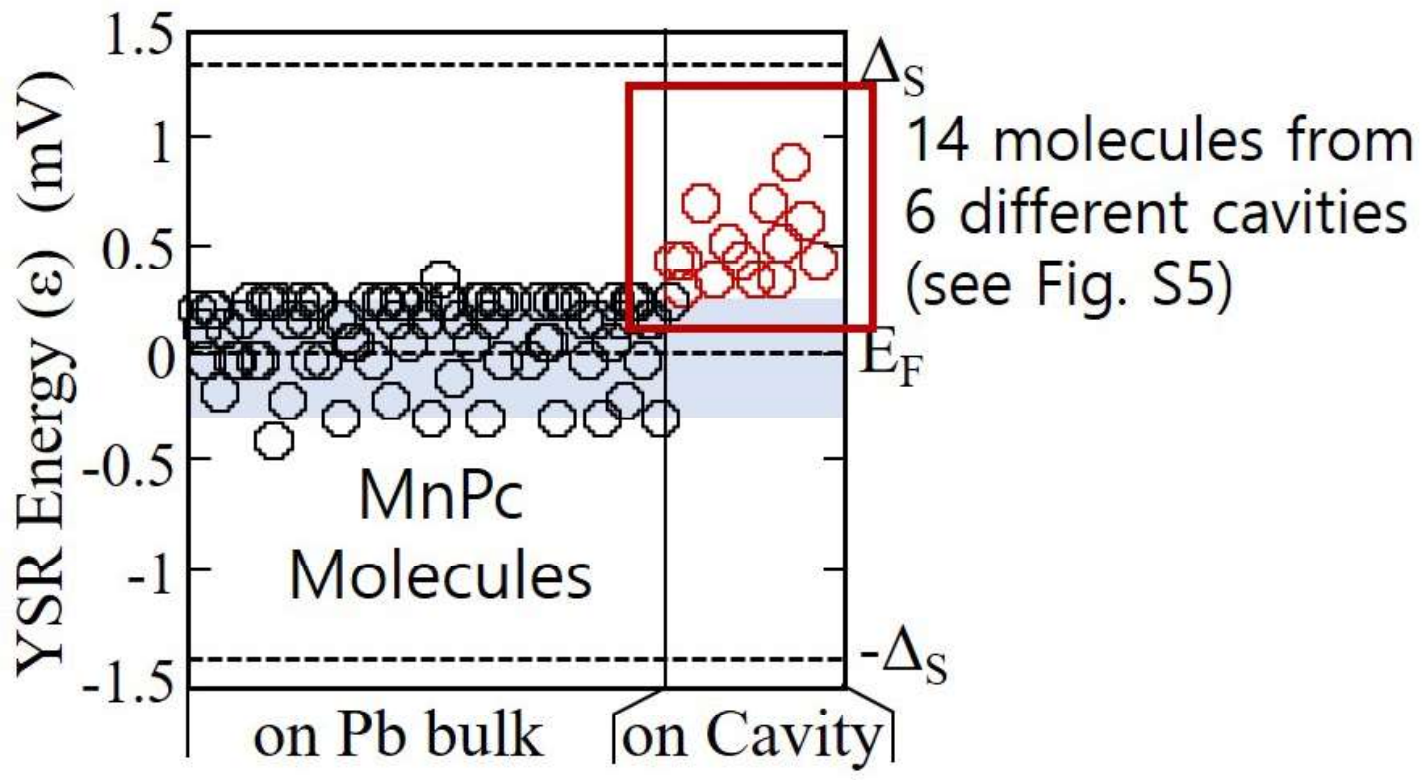

Figure S4. The YSR energies of 14 MnPc molecules on 6 different Ar cavities. The details of the cavities are provided in Figure S5.

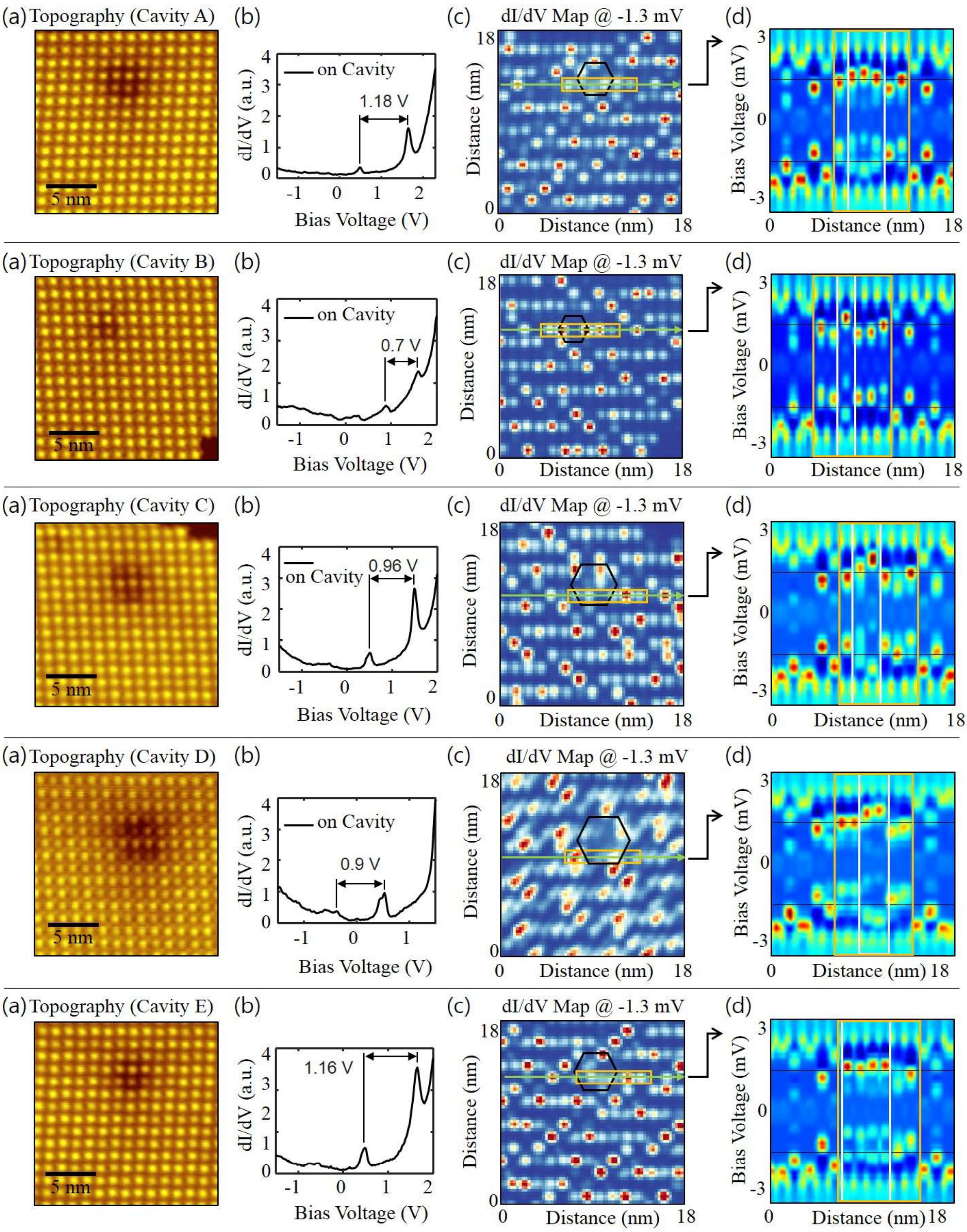

(Continue on the next page)

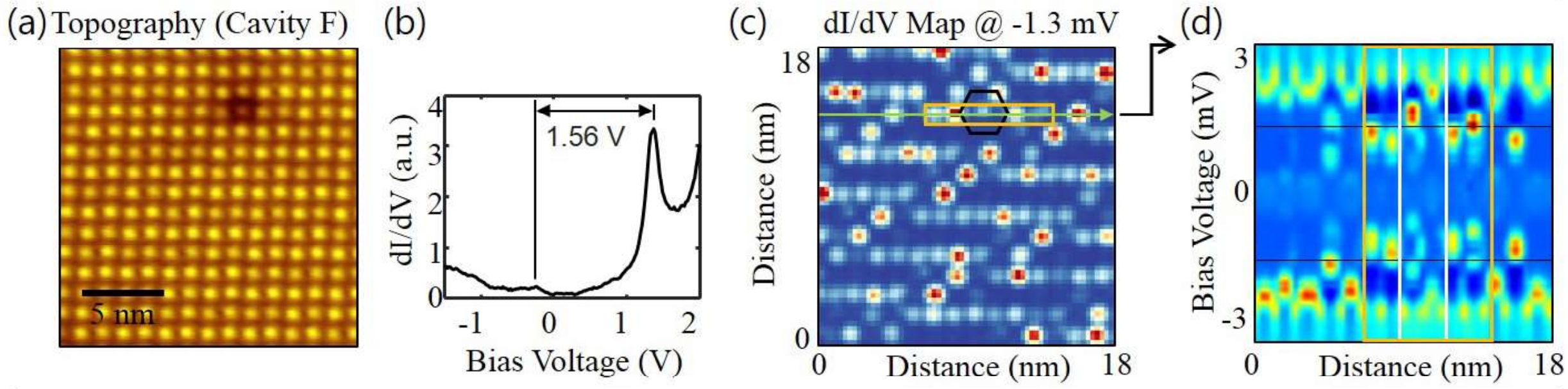

Figure S5. Each data set displays (a) topographic image at $V_{bias}$ = 100 mV and I = 50 pA. The dark contrast of the molecules in the cavity ensures that DOS is suppressed at $E_F$ by the cavity. (b) $dI/dV$ spectrum showing the quantum well oscillations. (c) $dI/dV$ map at -1.3 mV. (d) $dI/dV$ line spectra taken along the green arrow in the $dI/dV$ map. The white box in (d) indicates where the cavity is located.

## 3 Kondo Resonance of Co Atoms on PbSe/Pb(111)

To check if the Co atoms on 1-ML PbSe/Pb(111) show the Kondo resonance, we raised the experiment temperature to 7.7 K. Figure S6 displays the spectra measured on/off the Co atom. The Kondo resonance is clearly resolved on the Co atom, revealing that the exchange coupling between the magnetic impurities and conduction electrons is antiferromagnetic.

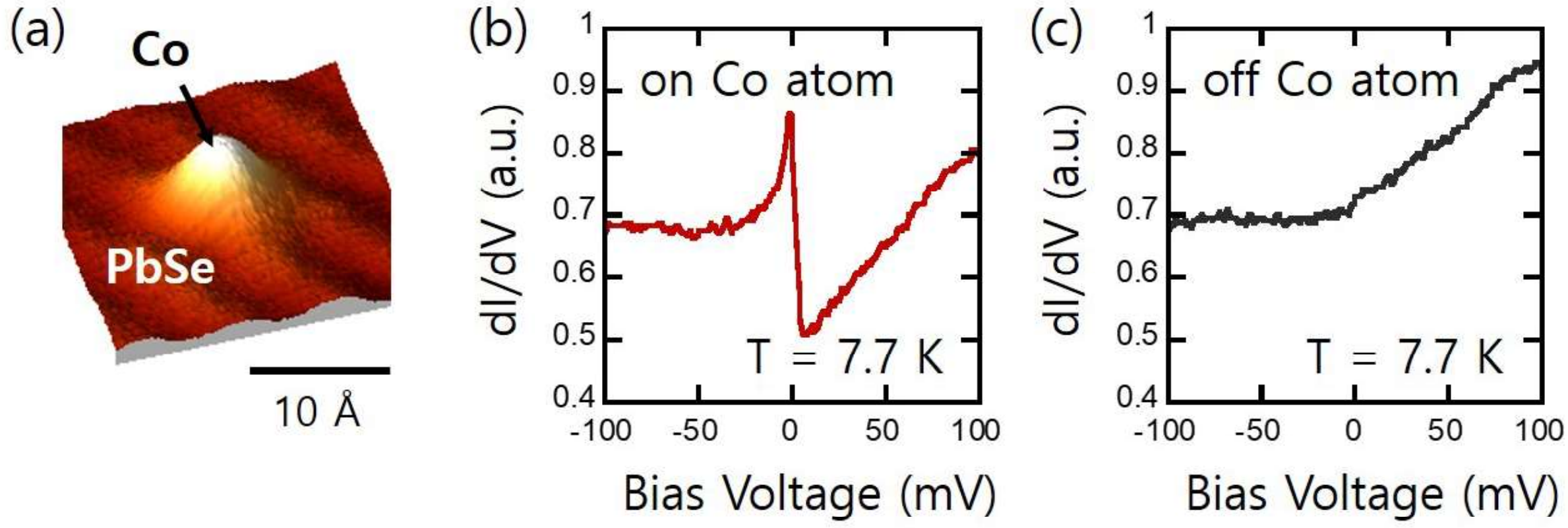

Figure S6. (a) Topography of the Co atom on PbSe/Pb(111). (b) $dI/dV$ spectrum measured on the Co atom. (c) $dI/dV$ spectrum measured on bare PbSe surface.

## 4 Theoretical Formulation of the YSR Bound States

### *4.1 Model Hamiltonian of a Magnetic Impurity on a Superconductor*

The Bogoliubov-de Gennes (BdG) Hamiltonian ($H_{SC,4\times4}$) of an s-wave superconductor and a point-like impurity potential ($V_{imp,4\times4}\delta^{(d)}(r)$) is given by

$$H_{SC,4\times4} = \xi_p \tau_z I_{2\times2} + \Delta_s \tau_x I_{2\times2} \tag{1}$$

$$V_{imp,4\times4}\delta^{(d)}(r) = U_{imp}\tau_z I_{2\times2}\delta^{(d)}(r) + J_{imp} I_{2\times2} S_{imp} \cdot \sigma \delta^{(d)}(r). \tag{2}$$

Here $p$ and $r$ denote the momentum and the position of the electron. $\xi_p = \frac{p^2}{2m} - \mu$ is the electron energy with the chemical potential $\mu$, $\Delta_s$ is the superconducting gap, and $d$ is the dimension of the system. $J_{imp}$ represents the exchange coupling between the magnetic impurity $S_{imp}$ and the electron in the superconductor. $\tau(\sigma)$ are the Pauli matrices in Nambu (spin) space. The basis of BdG Hamiltonian is chosen to be four-component Nambu operator $\psi = \begin{pmatrix} c_{p,\uparrow} & c_{p,\downarrow} & c^{\dagger}_{-p,\downarrow} & -c^{\dagger}_{-p,\uparrow} \end{pmatrix}^T$ with the electron field operators $c^{\dagger}_{p,\sigma}$ and $c_{p,\sigma}$. We choose $S_{imp}$ aligned along the z direction in the calculation.

The Schrödinger equation for the magnetic impurity on the s-wave superconductor is written by

$$(EI_{2\times2}I_{2\times2} - H_{SC,4\times4})\psi(r) = V_{imp,4\times4}\psi(0). \tag{3}$$

The YSR bound states at $r$ = 0 will be represented by $|\psi(0)|^2$. By multiplying both sides of Eq. (3) with $(EI_{2\times2}I_{2\times2} - H_{SC,4\times4})^{-1}(= G^0(E, r, 0))$, the spatial oscillation of the wavefunction can be obtained

$$\psi(r) = G^0(E, r, 0)V_{imp,4\times4}\psi(0). \tag{4}$$

Here $G^0$ is a bare Green's function of the superconductor. We introduce the linearlization of momentum around Fermi energy with $p(\xi) = p_F + \frac{\xi}{v_F}$,[1,2] where $p_F$ is the Fermi wavenumber and $v_F$ is the Fermi velocity. By extending the cut-off to $[-\infty, \infty]$,

$$G^0(E, r, 0) \simeq g_{dD,F} \int \frac{d\Omega_{d-1}}{(2\pi)^{d-1}} \int_{-\infty}^{\infty} d\xi e^{ir(p_F + \frac{\xi}{v_F})\cos\theta} \frac{(EI_{2\times2} + \xi\tau_z + \Delta_s\tau_x)I_{2\times2}}{E^2 - \Delta_s^2 - \xi^2}, \tag{5}$$

where $g_{dD,F}$ is the density of states (DOS) of normal state at the Fermi energy.

To compute $G^0(E,r,0)$, we need to calculate following integrals. Considering 3-dimensional (3D) case,

$$I_A = \frac{g_{3D,F}}{2\pi}\int_{-1}^{1} d\cos\theta \int_{-\infty}^{\infty} d\xi e^{ir(p_F+\frac{\xi}{v_F})\cos\theta}\frac{1}{\xi^2+\Delta_s^2-E^2} \tag{6}$$

$$I_B = \frac{g_{3D,F}}{2\pi}\int_{-1}^{1} d\cos\theta \int_{-\infty}^{\infty} d\xi e^{ir(p_F+\frac{\xi}{v_F})\cos\theta}\frac{\xi}{\xi^2+\Delta_s^2-E^2}\frac{\omega_D^2}{\xi^2+\omega_D^2}. \tag{7}$$

Here we introduced a convergence factor $(\omega_D^2/(\xi^2+\omega_D^2))$ into $I_B$,[1] where the Debye frequency $\omega_D$ is an UV cut-off. Performing the integrals over $\xi$ and $cos\theta$ in a sequence, we find

$$I_A = \frac{\pi g_{3D,F}}{\sqrt{\Delta_s^2-E^2}}\cdot\frac{1}{2\pi}\left(\frac{2k_s}{p_F^2+k_s^2}\frac{1}{r}+\frac{-2p_F\sin(p_Fr)+2k_s\cos(p_Fr)}{p_F^2+k_s^2}\frac{1}{r}e^{-k_sr}\right) \tag{8}$$

$$\begin{aligned} I_B = {} & i\pi g_{3D,F}\cdot\frac{\omega_D^2}{\omega_D^2-\omega^2}\frac{1}{2\pi}\left(\frac{-2p_F}{p_F^2+k_s^2}\frac{i}{r}+\frac{2p_F\cos(p_Fr)+2k_s\sin(p_Fr)}{p_F^2+k_s^2}\frac{i}{r}e^{-k_sr}\right.\\ & \left.-\frac{-2p_F}{p_F^2+k_{(\frac{\omega_D}{v_F})}^2}\frac{i}{r}-\frac{2p_F\cos(p_Fr)+2(\frac{\omega_D}{v_F})\sin(p_Fr)}{p_F^2+(\frac{\omega_D}{v_F})^2}\frac{i}{r}e^{-\frac{\omega_D}{v_F}r}\right), \end{aligned} \tag{9}$$

where $k_s = \sqrt{\Delta_s^2-E^2}/v_F$. Taking the limit of $k_F \gg k_s, k_F \gg \frac{\omega_D r}{v_F} \gg 1$, they are reduced to

$$I_A \simeq -\frac{\pi g_{3D,F}}{\sqrt{\Delta_s^2-E^2}}\cdot\frac{1}{\pi}\frac{sin(p_Fr)}{p_Fr}e^{-k_sr} \tag{10}$$

$$I_B \simeq -\pi g_{3D,F}\cdot\frac{1}{\pi}\frac{cos(p_Fr)}{p_Fr}e^{-k_sr}. \tag{11}$$

The results are consistent with [1]. The matrix form of $G^0(E,r,0)$ is then

$$G^0(E,r,0) = \begin{pmatrix} EI_A+I_B & 0 & \Delta_sI_A & 0 \\ 0 & EI_A+I_B & 0 & \Delta_sI_A \\ \Delta_sI_A & 0 & EI_A-I_B & 0 \\ 0 & \Delta_sI_A & 0 & EI_A-I_B \end{pmatrix}. \tag{12}$$

Note that $I_A = -\frac{\pi g_{dD,F}}{\sqrt{\Delta_s^2-E^2}}\frac{\Omega_{d-1}}{(2\pi)^{d-1}}$ and $I_B = 0$ when $r = 0$, which is useful for calculating $G^0(E,r=0,0)$.

*4.2 The YSR Bound States as a Solution of the Model Hamiltonian*

The Schrödinger equation at the impurity site can be obtained by by setting $r = 0$ in Eq. (4),

$$(I_{2\times 2}I_{2\times 2} - G^0(E,0,0)V_{imp,4\times 4})\psi(0) = 0 \tag{13}$$

or

$$[I_{2\times 2}I_{2\times 2} + \frac{1}{\sqrt{\Delta_s^2 - E^2}}\begin{pmatrix} E\beta_+ & 0 & -\Delta_s\beta_- & 0 \\ 0 & E\beta_- & 0 & -\Delta_s\beta_+ \\ \Delta_s\beta_+ & 0 & -E\beta_- & 0 \\ 0 & \Delta_s\beta_- & 0 & E\beta_+ \end{pmatrix}]\psi(0) = 0 \tag{14}$$

Here we used $G^0(E,0,0)$, $V_{imp,4\times 4}$ and dimensionless parameters $\beta_\pm$.

$$G^0(E,0,0) = -\frac{\pi g_{dD,F}}{\sqrt{\Delta_s^2 - E^2}}\frac{\Omega_{d-1}}{(2\pi)^{d-1}}\begin{pmatrix} E & 0 & \Delta_s & 0 \\ 0 & E & 0 & \Delta_s \\ \Delta_s & 0 & E & 0 \\ 0 & \Delta_s & 0 & E \end{pmatrix}, \tag{15}$$

$$V_{imp,4\times 4} = \begin{pmatrix} U_{imp} + J_{imp} & 0 & 0 & 0 \\ 0 & U_{imp} - J_{imp} & 0 & 0 \\ 0 & 0 & -U_{imp} + J_{imp} & 0 \\ 0 & 0 & 0 & -U_{imp} - J_{imp} \end{pmatrix}, \tag{16}$$

$$\beta_\pm = (U_{imp} \pm J_{imp}S_{imp})\pi g_{3D,F}\frac{\Omega_{3-1}}{(2\pi)^{3-1}} = g_{3D,F}(U_{imp} \pm J_{imp}S_{imp}). \tag{17}$$

We used $\Omega_{3-1} = 4\pi$ in the 3D case. By diagonalizing Eq. (14) with $\delta^\pm = \arctan\beta_\pm$, where $\delta^\pm \in [-\pi/2, \pi/2]$, we obtain

$$\begin{cases} E_{\uparrow}^{electronic} & = -sgn(J_{imp})\cos(\delta^+ - \delta^-) \mid \Delta_s \mid \\ \psi_{\uparrow}(0) & = \frac{1}{\mathcal{N}} \begin{pmatrix} -sgn(\Delta_s)sgn(J_{imp})\cos\delta^+ \\ 0 \\ \cos\delta^- \\ 0 \end{pmatrix} \end{cases} \begin{cases} E_{\downarrow}^{electronic} & = sgn(J_{imp})\cos(\delta^+ - \delta^-) \mid \Delta_s \mid \\ \psi_{\downarrow}(0) & = \frac{1}{\mathcal{N}} \begin{pmatrix} 0 \\ sgn(\Delta_s)sgn(J_{imp})\cos\delta^- \\ 0 \\ \cos\delta^+ \end{pmatrix} \end{cases} \tag{18}$$

where $\mathcal{N}$ is the normalization factor. In the experiment, the first component of the spin-up wavefunction and the second component of the spin-down wavefunction contribute to the STM tunneling current because they are related to the particle parts of Nambu operator.[4] They constitute a pair of YSR bound states in the superconducting gap.

Next, we compute the spatial oscillations of the wavefunctions using Eq. (4). From the asymptotic forms of the integrals in Eqs. (10) and (11), we find

$$\psi_{\uparrow}(r) = \frac{1}{\mathcal{N}} \begin{pmatrix} -sgn(\Delta_s)sgn(J_{imp})\sin(p_F r - \delta^+) \\ 0 \\ \sin(p_F r - \delta^-) \\ 0 \end{pmatrix} \frac{1}{r} e^{-r|sin(\delta^+ - \delta^-)|/\xi} \tag{19}$$

$$\psi_{\downarrow}(r) = \frac{1}{\mathcal{N}} \begin{pmatrix} 0 \\ sgn(\Delta_s)sgn(J_{imp})\sin(p_F r - \delta^-) \\ 0 \\ \sin(p_F r - \delta^+) \end{pmatrix} \frac{1}{r} e^{-r|sin(\delta^+ - \delta^-)|/\xi}, \tag{20}$$

where $\xi = v_F/\Delta_S$ is the coherence length of the superconductor at T = 0 K.[3]

## 5 Spatial Oscillations of YSR Bound States of the Co Atom on PbSe/Pb(111)

The line profiles of the oscillating YSR bound states in Figure 4g (main text) are obtained along the arrows depicted in Figure S7(a). We obtained better signal-to-noise ratio when taking the line profiles than the radial averages (Figure S7(b)). Nevertheless, the position of the oscillation peaks coincides with each other for the line profiles and the radial averages (Figure S7(c)).

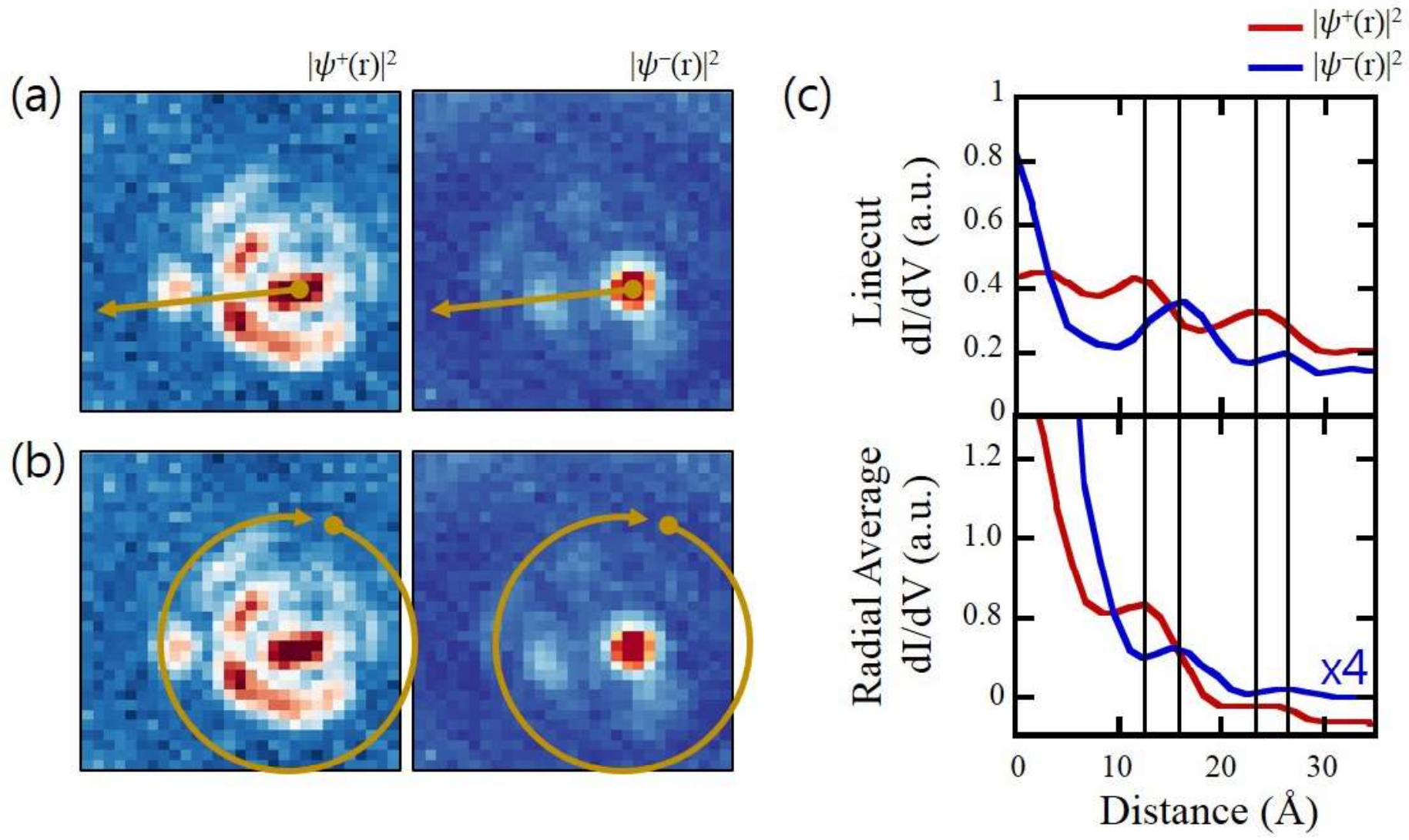

Figure S7. Comparison between line profiles and radial averages of the electron-like and hole-like YSR bound states.

## 6 Determination of Sizes of Sample- and Tip- Superconducting Gaps

The superconducting gap of the sample ($\Delta_S$) and the tip ($\Delta_T$) can be calculated using the thermal replica states of the YSR bound states. Figure S8(a) shows the energy position of peaks when a superconducting tip is used to measure the YSR bound states of sample. Schematic diagram of tunneling corresponding to each peak is provided in Figure S8(d). The YSR bound state at the negative bias voltage appears at $-(\Delta_T + \epsilon) = -1.9$ mV and its replica state appears at $(\Delta_T - \epsilon) = 0.3$ mV (Figure S8(a) and S8(b)). We obtain $\Delta_T = 1.1$ mV by solving these linear equations. From the position of coherent peak, $\Delta_T + \Delta_S = 2.4$ mV, we obtain $\Delta_S = 1.3$ mV. The YSR energy is calculated by $-(\Delta_T+\epsilon)-(-\Delta_T) = (-1.9 \text{ mV}) - (-1.1 \text{ mV}) = -0.8$ mV (Figure S8(c) and Figure 4(d) in main text). The sizes of superconducting gaps were similar for other different tips. The tip superconducting gap was typically smaller than the sample superconducting gap most likely due to the relatively small amount of Pb coated on the tip.

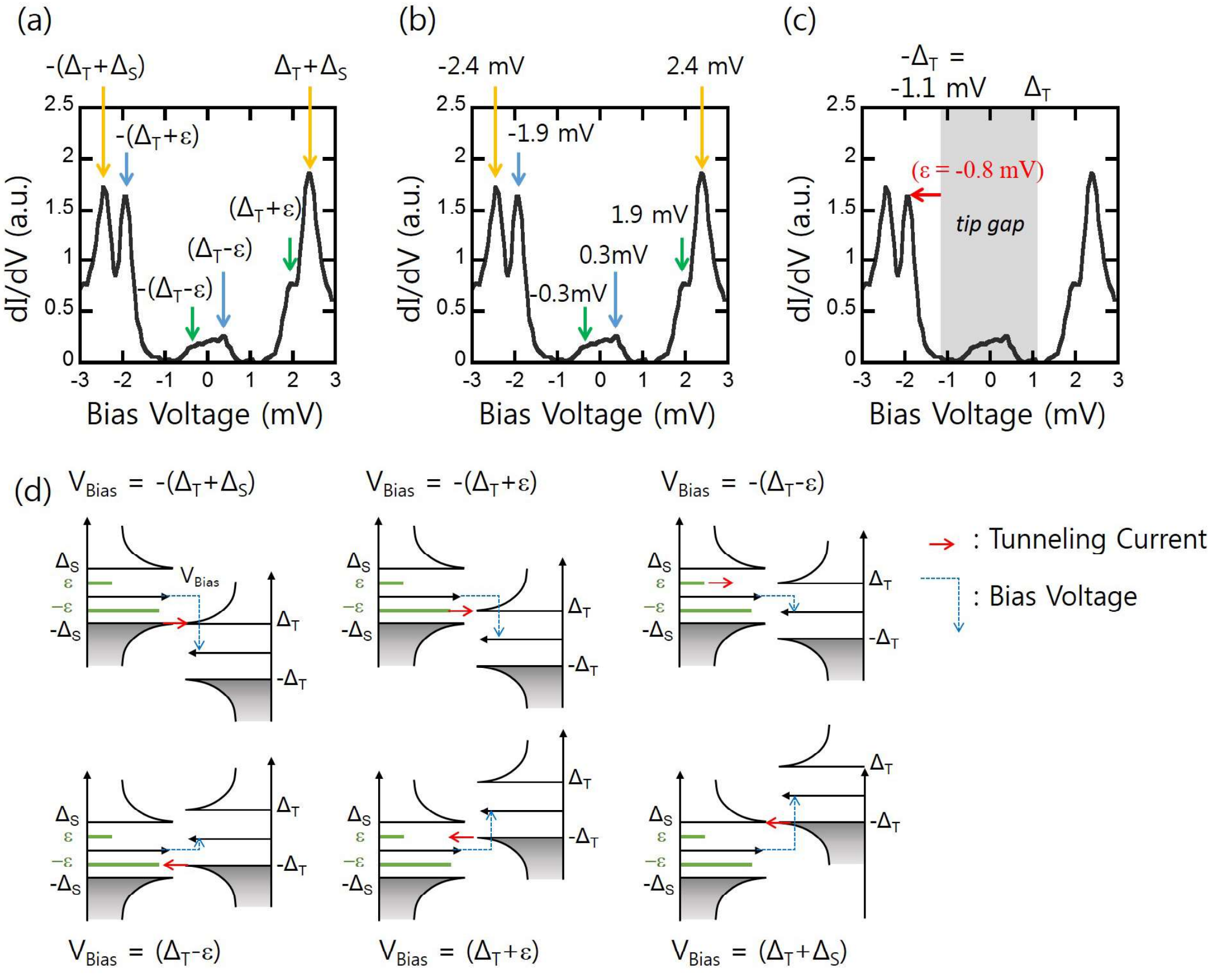

Figure S8. The sample superconducting gap and the tip superconducting gap can be calculated using the thermal replica states of YSR bound states.

## 7 Assignment of Calculated YSR Bound States to the Measured YSR Bound States

In the main text, we have shown the sign of scattering potential $U$ can be unambiguously determined by examining the trend of the YSR energy shift as $\rho$ decreases ($i.e.$, $U > 0$ for MnPc molecules on Pb(111) and $U < 0$ for Co atoms on PbSe/Pb(111)). Once the sign of $U$ is determined, it is easy to judge whether a system is in the strong coupling regime or in the weak coupling regime from a single spectrum of YSR bound states. When the YSR bound state at the negative bias voltage ($|\psi^-|^2$) is stronger in amplitude than that at the positive bias voltage ($|\psi^+|^2$), it is in the weak coupling regime if $U < 0$ and in the strong coupling regime if $U > 0$. From Eqs. (18) and (19), the energies of the spin-up part and spin-down part of the YSR bound states and their corresponding wavefunctions in space are given by

$$\begin{cases} E_\uparrow = -sgn(J_{imp}) \cos(\delta^+ - \delta^-) \mid \Delta_s \mid \\ \psi_\uparrow(r) \propto \dfrac{sin(p_F r - \delta^+)}{r} e^{-r|sin(\delta^+ - \delta^-)|/\xi} \end{cases} \qquad \begin{cases} E_\downarrow = sgn(J_{imp}) \cos(\delta^+ - \delta^-) \mid \Delta_s \mid \\ \psi_\downarrow(r) \propto \dfrac{sin(p_F r - \delta^-)}{r} e^{-r|sin(\delta^+ - \delta^-)|/\xi} \end{cases} \tag{21}$$

We will match the calculated YSR bound states ($\psi_\uparrow, \psi_\downarrow$) to the measured YSR bound states ($\psi^+$, $\psi^-$). The signs of $E_\uparrow$ and $E_\downarrow$ depend on (a) sign of $J_{imp}$ and (b) if $\delta^+ - \delta^- > \pi/2$ (strong coupling regime) or $\delta^+ - \delta^- < \pi/2$ (weak coupling regime). We can determine both conditions experimentally. For example, the spectrum in Figure 4d (main text) indicates that the system is in the weak coupling regime since the amplitude of $|\psi^-|^2$ is stronger than that of $|\psi^+|^2$ under $U < 0$. We determined $J_{imp} > 0$ for the Co atoms on PbSe/Pb(111) from the Kondo measurement. Therefore, from Eq. (21), $E_\uparrow$ locates on the negative energy and $E_\downarrow$ locates on the positive energy in the spectrum. This enables us to conclude that $\psi^+$ and $\psi^-$ are associated with $\psi_\downarrow$ and $\psi_\uparrow$, respectively. Therefore, $\psi^+ \propto sin(p_F r - \delta^-)e^{-r|sin(\delta^+ - \delta^-)|/\xi}/r$ and $\psi^- \propto sin(p_F r - \delta^+)e^{-r|sin(\delta^+ - \delta^-)|/\xi}/r$.

## 8 Simulation of YSR Bound States of the Co Atom on PbSe/Pb(111)

The theory shows that the wavefunctions of YSR bound states are given by

$$\psi_\uparrow(r), \psi_\downarrow(r) \propto \frac{sin(k_F r - \delta^\pm)}{r} exp(-r|sin(\delta^+ - \delta^-)|/\xi), \tag{22}$$

where $\xi$ is the coherence length of the superconductor. $\delta^+$ and $\delta^-$ are the scattering phases for spin-up and spin-down parts of YSR bound states, respectively. In the weak coupling regime and $J_{imp} > 0$, the phase shift $\delta^+$ is associated with the YSR bound state at the negative bias voltage ($|\psi^-|^2$) and $\delta^-$ is related to the YSR bound state at the positive bias voltage ($|\psi^+|^2$). We simulated the YSR oscillations using Eq. (21) and compared the results with experimental data (Figure 4 in main text). In the simulation process, we fixed $k_F$ = 2.9 nm$^{-1}$ and $\delta^+ - \delta^-$ = 0.3$\pi$ which were extracted from Figure 4g in main text. The coherence length $\xi$ = 83 $nm$ is used for Pb. Under the given constraints, we have found the simulation results with $\delta^+ = -0.15\pi$ and $\delta^- = -0.45\pi$ agree well with the experimental data (Figure S9). These values of phase shifts are self-consistent with $J_{imp} > 0$ and $U < 0$ for the Co atom on PbSe/Pb(111).

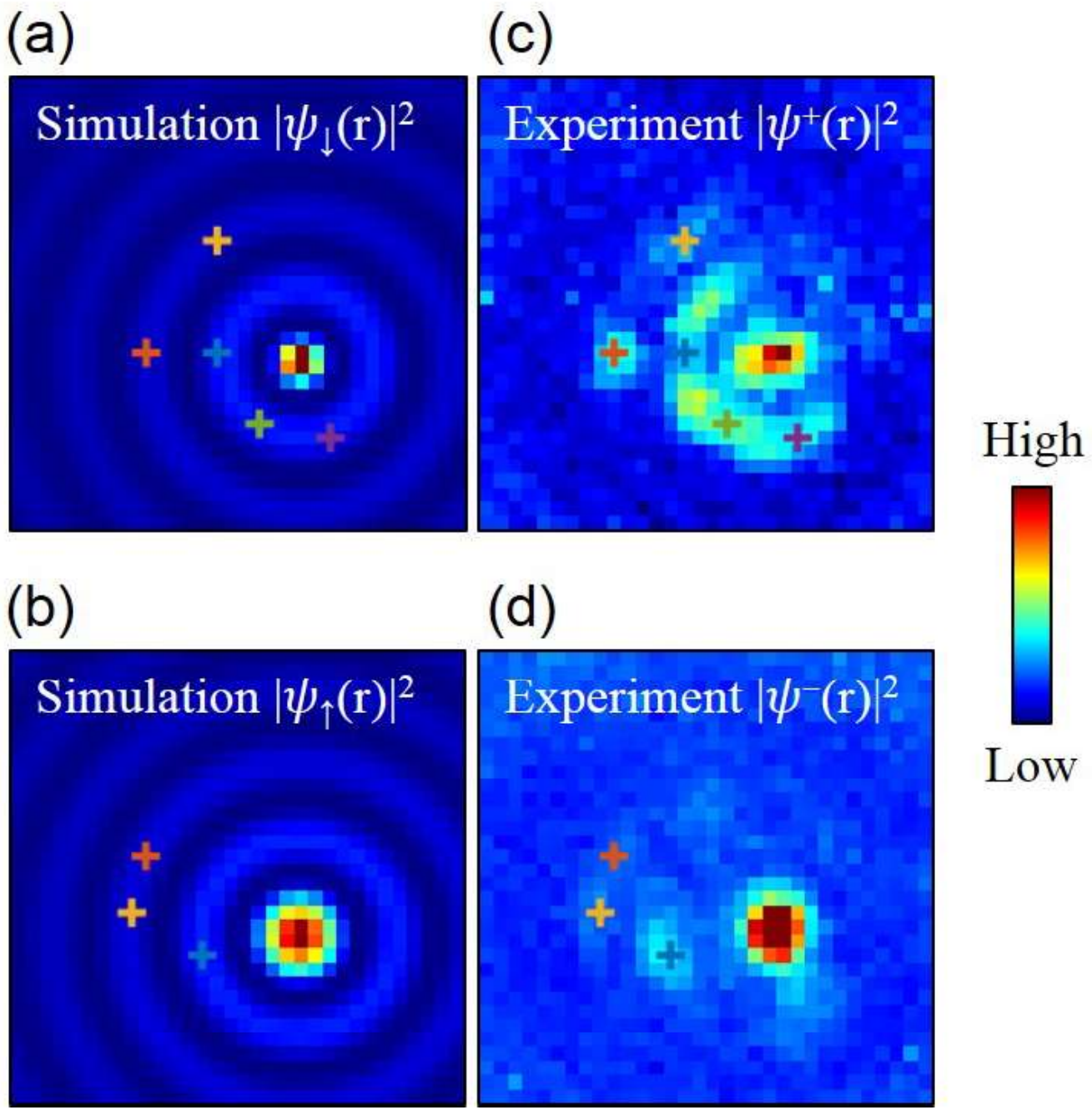

Figure S9. (a,b) Simulation of the electron-like and hole-like parts of the YSR bound states. The parameter values of $k_F$ = 2.9 nm$^{-1}$, $\delta^- = -0.45\pi$ ($|\psi^+(r)|^2$) and $\delta^+ = -0.15\pi$ ($|\psi^-(r)|^2$) are used for the simulation. (c,d) The experiment data. The image size is 6.6 nm by 6.6 nm. The cross marks are given for an eye-guide for comparing between the simulation and the experiment.

## 9 Experimental Parameters for Figures in Main Text

| Figure Name | Bias Voltage ($V_{bias}$) | Tunneling Current ($I_t$) | Lock-in Frequency ($f$) | Lock-in Amplitude ($V_{mod}$) |
|---|---|---|---|---|
| Figure 1(a) | -0.1 V | 50 pA | N/R (non-relevant) | N/R |
| Figure 2(a) | -0.1 V | 50 pA | N/R | N/R |
| Figure 2(b) | -0.7 V | 100 pA | 463 Hz | 10 mV |
| Figure 2(c) | -3 mV | 100 pA | 463 Hz | 60 $\mu$V |
| Figure 2(e) | -3.2 mV | 100 pA | 463 Hz | 60 $\mu$V |
| Figure 2(f) | -3.2 mV | 100 pA | 463 Hz | 60 $\mu$V |
| Figure 3(a) | -0.1 V | 50 pA | N/R | N/R |
| Figure 3(b) | -1 V | 100 pA | 463 Hz | 10 mV |
| Figure 3(c) | -3 mV | 100 pA | 463 Hz | 60 $\mu$V |
| Figure 3(d) | -3 mV | 100 pA | 463 Hz | 60 $\mu$V |
| Figure 4(a) | -0.1 V | 50 pA | N/R | N/R |
| Figure 4(b) | -1 V | 100 pA | 463 Hz | 10 mV |
| Figure 4(c) | -3 mV | 100 pA | 463 Hz | 60 $\mu$V |
| Figure 4(d) | -3 mV | 100 pA | 463 Hz | 60 $\mu$V |

Table 1. STM Parameters for Figures in Main Text